\documentstyle[aps,epsf,prl,preprint]{revtex}
\draft
\preprint{Accepted as Rapid Comm in PRE}
\begin{document}

\title{Competition between Spiral-Defect Chaos and Rolls in Rayleigh-B\'enard
Convection} \author{Kapil M. S. Bajaj, David S. Cannell, and Guenter Ahlers}
\address{Department of Physics and Center for Nonlinear Science, University
of California, Santa Barbara, California 93106, USA} \date{\today} \maketitle
\begin{abstract}
We present experimental results for pattern formation in Rayleigh-B\'enard
   convection of a fluid with a Prandtl number $\sigma \simeq 4$. We find
   that the spiral-defect-chaos (SDC) attractor which exists for $\sigma
   \simeq 1$ has become unstable. Gradually increasing the temperature
   difference $\Delta T$ from below to well above its critical value $\Delta
   T_c$ no longer leads to SDC. A sudden jump of $\Delta T$ from below to
   above $\Delta T_c$ causes convection to grow from thermal fluctuations and
   does yield SDC. However, the SDC is a transient; it coarsens and forms a
   single cell-filling spiral which then drifts toward the cell wall and
   disappears.  
\end{abstract}
\vskip 0.3in 

Rayleigh-B\'enard convection (RBC) occurs in a shallow horizontal layer of a
fluid heated from below when the temperature difference $\Delta T$ across the
layer exceeds a critical value $\Delta T_c$. The velocity fields which form
above $\Delta T_c$ have long been used as an experimental testing ground for
theories of pattern formation in nonlinear systems.~\cite{CH93} The phenomena
which occur depend upon the Prandtl number $\sigma = \nu / \kappa$, the
dimensionless ratio of the kinematic viscosity $\nu$ to the thermal
diffusivity $\kappa$.  A fascinating recent discovery in RBC is that of
spiral-defect chaos (SDC) for $\sigma \simeq 1$.~\cite{MBCA93} It consists of
patches of rotating convection-roll spirals and of other defects, and occurs
for $\epsilon \equiv \Delta T / \Delta T_c - 1 > \epsilon_s > 0$. The spirals
are coherent structures with a typical diameter of a few roll
wavelengths. They can be right- or left-handed, single- or multi-armed, and
appear and disappear irregularly in time and space. SDC has since been found
in numerical solutions of model equations~\cite{Models} and of the
Navier-Stokes equations (NSE)~\cite{DPW94}. The band of wavenumbers for
SDC~\cite{MBCA93,DPW94} lies well within the range for which straight rolls
are also stable~\cite{Busse} in the laterally infinite system. Thus the
solutions of the equations of motion of RBC exhibit bistability of
time-independent straight rolls and of SDC, with each of the two having
distinct attractor basins. For $\sigma \alt 1$ the generic initial conditions
of real RBC experiments lie in the attractor basin of SDC.\cite{MBCA93,LA96}
Here we report that this situation is changed dramatically when $\sigma$ is
increased.  For $\sigma = 4.0 \pm 0.2$, we do not find any spirals when
increasing $\epsilon$ in small steps from below zero to large positive
values. Thus, this experimental path does not cross an attractor basin of
SDC. However, when $\epsilon$ is increased suddenly from below zero to values
of ${\cal O}(1)$, SDC evolves in the sample interior from random
noise~\cite{WAC95}. Thus, random initial conditions appear to remain within
the attractor basin of SDC even for $\sigma \simeq 4$. But this is illusory
since the SDC turns out to be a transient rather than a stable attractor. At
very long times, the spirals coarsen, forming larger and larger spirals until
only one giant spiral fills nearly the entire cell.  At even longer times the
giant spiral is not stable either; its head slowly drifts toward the sidewall
and disappears from the cell, leaving a state similar to those obtained by
increasing $\epsilon$ gradually. Our observations differ from those of
Assenheimer and Steinberg,\cite{AS94,AS96} who reported SDC in a sample of
SF$_6$ near its critical point with Prandtl numbers similar to and even
larger than ours.

A detailed description of the apparatus was given in Ref.~\cite{LCA95}. The
convection cell had a 0.95 cm thick sapphire top plate and a diamond-machined
aluminum bottom plate.  A film heater glued to the lower side of the bottom
plate provided the heat current. The top of the sapphire plate was in contact
with a temperature-regulated circulating water bath. The temperatures for
both bottom and top plates were regulated to better than 1~mK. The circular
sidewall was made of high-density polyethylene. It had an inner diameter of
8.89 cm, and was sealed to the top and bottom plates with ethylene-propylene
O-rings.  High-purity liquid-chromatography grade acetone was dried by using
3\AA~molecular sieves followed by distillation, and then injected into the
cell with a syringe without significant exposure to the atmosphere. Two cells
were used. The cell height was $d = 0.80\pm 0.01 ~ (1.0\pm 0.01)$ mm,
corresponding to an aspect ratio (radius/height) $\Gamma \approx 55 ~
(45)$. The calculated value of $\Delta T_c$ was $8.56 ~ (4.50)$ $^\circ$C. At
onset, the coefficient $Q$ \cite{Bu67} describing the departure from the
Boussinesq approximation was $0.37 ~ (0.20)$. The vertical thermal diffusion
time $t_v$ was $6.8 ~ (10.5)$ s. The onset of convection was determined from
measurements of the Nusselt number $\cal N$ (the ratio of the effective
thermal conductivity to that in the conduction state). Quasi-static results
for ${\cal N}$ over the entire range of the experiment are shown in
Fig.~\ref{Nusselt}. Images of convection patterns were taken using
shadowgraph apparatus described previously~\cite{BBMTHCA96}.
	
In one set of experiments, $\epsilon$ was varied slowly from below zero to
above 5 by changing it in steps of about 0.1. The system was allowed to
equilibrate for at least one horizontal diffusion time $\Gamma^2 t_v$ (about
6 hours) after each step. This procedure corresponds to an effective ramp
rate $\beta \equiv d\epsilon / dt \leq 5.4 \times 10^{-5}$ (here $t$ is
measured in units of $t_v$). It yielded images like those shown in
Fig.~\ref{updownhex}. For $\epsilon \alt 1$, the patterns consisted
essentially of straight rolls, except for typically one or two grain
boundaries and perhaps one focus singularity adjacent to the sidewall. A
representative example is shown in Fig.~\ref{updownhex}a, which is for
$\epsilon = 0.5$. The wide range of $\epsilon$ over which relatively straight
rolls occur is in contrast to the $\sigma \simeq 1$ case, where significant
roll curvature occurs already for $\epsilon \agt 0.1$
\cite{Cr89,HEA95b}. Figures~\ref{updownhex}b, c, and d are for $\epsilon =$
2.0, 3.7, and 4.5. They show that the rolls do acquire significant curvature
as $\epsilon$ is increased. This curvature is associated with a tendency for
the roll axes to terminate perpendicular to the wall. It leads to the
formation of additional wall foci and to a reduction of the roll wavelength
in the cell center. For $\sigma \simeq 1$, this phenomenon occurred already
at $\epsilon$ values which were about an order of magnitude smaller, and
yielded a time-dependent state with repeated formation of defects in the cell
interior via the skewed-varicose mechanism.\cite{Cr89,HEA95b} In the present
case, the images were essentially independent of time at constant
$\epsilon$. No defect formation occurred because even the significantly
enhanced wavenumber range encountered in the cell interior did not result in
any crossing of the skewed-varicose instability-boundary. We illustrate this
latter point in Fig. \ref{balloon}, which at small wavenumbers gives the
Eckhaus and at large the skewed-varicose instability as a solid line. The
solid circles are the mean wavenumbers $\bar k$ from the present run. They
were obtained from the structure functions $S(k)$ (squares of the moduli of
the Fourier transforms) of the images as described
elsewhere~\cite{MBCA93}. The horizontal bars represent the widths of the
structure functions, obtained from the second moments of $S(k)$ about $\bar
k$. It is apparent from these data that even the large- and small-$k$ tails
of $S(k)$ are primarily well within the range of stable straight rolls of the
infinite system.

When $\epsilon$ was increased further, the formation of cellular flow began
near one of the wall foci, as shown in Fig.~\ref{updownhex}e for $\epsilon =
5.0$. During many horizontal diffusion times these flow cells multiplied and
spread over a wider part of the sample and evolved into a pattern of
coexisting regions of hexagons with either upflow or downflow at their
centers. These regions occurred together with curved rolls and wall foci in
other parts of the sample.  This is shown in Fig.~\ref{updownhex}f, which is
also for $\epsilon = 5.0$ but 6000$t_v$ after $\epsilon$ was last
increased. This pattern is similar to ones observed by Assenheimer and
Steinberg~\cite{AS96} (AS) with SF$_6$ close to its gas-liquid critical point
in a cell with $\Gamma = 80$ at $\epsilon=3.4$ and $\sigma=4.5$. However, the
threshold for and mechanism of hexagon generation seems different in our
case, since AS report that their hexagons appear already near $\epsilon =
2.8$, and via a core instability of spirals or targets which do not exist in
our system. Apparently the threshold for hexagon generation depends on the
nature of the previously existing structure.

Of greatest relevance to the present issue is that none of the images
obtained in runs like the one described above revealed any spirals. This is
so also for the larger $\Gamma = 55$. In contrast to this, for $\sigma \simeq
1$ and similar values of $\Gamma$\cite{HEA95b,LA96} the onset of SDC was
found at $\epsilon _s \simeq 0.55$, and SDC persisted to above $\epsilon
\simeq 3$ where the oscillatory instability was encountered at that $\sigma$
value. The total absence of spirals in our experiment also differs from the
results reported by AS, who for $\sigma = 4.3$ found spirals and targets
(zero-armed spirals) for $\epsilon \agt 0.84$.

In a second set of experiments we increased $\epsilon$ suddenly from
$\epsilon_0 = -1$ to $\epsilon_1 > 0$.  Care was taken to increase $\epsilon$
as quickly as possible to the final value in a smooth, monotonic manner
without overshooting. This procedure yielded small-amplitude random
convective flows in the cell center after only a few $t_v$. Examples are
shown in Fig.~\ref{spirals}a and e. The dynamic sidewall forcing during the
transient typically led to a small number of concentric rolls surrounding the
random flow field in the center.  Stabilizing at various values of
$\epsilon_1 \agt 1$, we observed a rapid growth (typically within $30 t_v$)
of targets and spirals from the disordered flow field, as shown in
Fig.~\ref{spirals}b and f. Simultaneously the mean wavevector decreased and
after a relatively short time settled down at the points shown by open
circles in Fig.~\ref{balloon}. Once the pattern of spirals, targets, and
other defects was generated, the complicated dynamics of SDC known from the
experiments at smaller $\sigma$ came into play. For smaller $\epsilon_1$ (say
$\epsilon_1 \simeq 0.5$), no spirals or targets developed and a pattern like
that of Fig.~\ref{updownhex}a evolved after a relatively short time.

In contrast to the experiments with $\sigma \simeq 1$, the SDC state which
for $\epsilon_1 \agt 1$ grew from the random initial conditions was not
stable. Instead of persistent SDC dynamics, we observed a coarsening of the
patterns over time. This is illustrated by Figs.~\ref{spirals}c and g. The
spirals grew in size until after a few horizontal diffusion times there was
left primarily one giant, nearly cell-filling, slowly rotating spiral as
shown in Figs.~\ref{spirals}d and h.

Although the giant spirals lived for several horizontal diffusion times,
experiments over even longer times showed that they were not really
stable. This is illustrated in Fig.~\ref{long_time}. At $7500t_v$ after a
jump in $\epsilon$ to $\epsilon_1 = 1.6$, one giant spiral is positioned with
its core centered in the cell. As shown by the other images, this core
location is not stable; the core gradually drifts towards the wall while the
radius of the spiral decreases so that it roughly equals the distance of the
core from the wall. During this process, no new spirals or targets are
born. At a time shortly after that of image d, the core left the cell
altogether and a state of straight/curved rolls and wall foci similar to the
ones obtained by the slow increase of $\epsilon$ (Fig.~\ref{updownhex})
remained.

For $\sigma \simeq 4$ and $\Gamma = 45$ or 55, we found that SDC can be
attained from the random flow fields generated by a jump in $\epsilon$, but
not by quasi-static changes of $\epsilon$. This is in contrast to $\sigma
\alt 1$ where SDC is difficult to avoid for $\Gamma \agt 30$
\cite{MBCA93,LA96} no matter how $\epsilon$ is varied.  We also demonstrated
by experiments extending over very long times that SDC, once formed, is
unstable in our samples with $\sigma \simeq 4$. It coarsens and evolves into
a single cell-filling spiral, whose core then drifts from the cell center
towards the sidewall where it disappears. Our results differ from those of
Assenheimer and Steinberg~\cite{AS94,AS96}, who found SDC without coarsening 
over the wide range
$2.8 \alt \sigma \alt 28$, with the spirals gradually being replaced by
targets as $\sigma$ increases. The reason for this difference is
unclear. However, we note several differences between the two experiments. In
one run with $\sigma = 4.5$ described by AS in some detail\cite{AS96} they
used a cell for which $\Delta T_c$ was two orders of magnitude smaller than
ours. Thus, assuming that their temperature control was similar to ours,
their experimental noise (spatial as well as temporal) in $\epsilon$
presumably was two orders of magnitude larger than ours. 
Furthermore extremely slow ramp rates $\beta \simeq 5\times 10^{-5}$
like ours would be more difficult to attain when $\Delta T_c$ is very
small. Also, the total time intervals over which they observed SDC at 
constant $\epsilon$ were  only ${\cal O}(\Gamma^2\tau_v)$~\cite{ASthesis} 
and are at least an order of magnitude smaller than ours 
($\approx 18\Gamma^2 t_v$). 
Finally we note that AS used a sample with $\Gamma = 80$ which is a 
factor of 1.5 larger than one of ours. One might conjecture that larger 
$\Gamma$ ($=80$, the AS value) stablizes SDC for $\sigma \simeq 4$ whereas
$\Gamma=55$ (our larger value) destabalizes it.
This would require a strong $\sigma$ dependence of
$\epsilon_s(\sigma,\Gamma)$ because SDC is readily
attained for instance for $\Gamma = 30$ and $\sigma = 1$. 
If this trend in  $\sigma$ dependence continues for $\sigma > 4$, 
AS should not have observed SDC in their cell with $\Gamma = 80$ at their 
much higher Prandtl numbers, up to 20 or larger. We conclude that the 
differences between the two experiments remain unresolved, and that 
additional measurements, particularly as a
function of $\Gamma$, are required to shed light on this
problem. Unfortunately it is difficult to reach larger $\Gamma$ values using
acetone because $\Delta T_c$ becomes very large if $d$ is reduced.

Finally we remark that it would be interesting to search for the coarsening
and instability of SDC in solutions of the Navier-Stokes equations.

We are grateful to Werner Pesch for permitting us to use
his program and for his advice during the calculation of the stability
boundary in Fig.~\ref{balloon}. This work was supported by the National
Science Foundation through Grant DMR94-19168.


\begin{figure}[p]
\caption{The Nusselt number ${\cal N}$ vs. $\epsilon$ for $\Gamma = 45$
obtained while increasing $\epsilon$ quasistatically. For this run, the top
temperature was 16.5$^\circ$C and $\Delta T_c = 4.50$ $^\circ$C.}
\label{Nusselt}
\end{figure}

\begin{figure}[p]
\caption{Representative images obtained during the experimental run of
Fig.~\protect{\ref{Nusselt}} where $\epsilon$ was increased gradually. a)
$\epsilon = 0.5$; b) $\epsilon = 2.0 $; c) $\epsilon = 3.7$; d) $\epsilon =
4.5$. Up to this value of $\epsilon$, ``rolls'' (with curvature, wall foci,
etc.) are stable and time independent. Images e) and f) are for $\epsilon =
5.0$, 1000$t_v$ and 6000$t_v$ after increasing $\epsilon$. They show the
co-existence of hexagon domains with upflow and downflow at the cell centers
and with rolls and wall foci. No spirals or targets were encountered.}
\label{updownhex}
\end{figure}

\begin{figure}[p]
\caption{The solid circles are the mean wavevectors obtained in the run of
Fig.~\protect{\ref{Nusselt}} by increasing $\epsilon$ quasistatically for
$\Gamma = 45$. The open circles are wavevectors obtained from the SDC state
before the coarsening process had progressed significantly. The horizontal
bars through the data points correspond to the width of the structure
function $S(k)$. The plus and cross are the wavevectors reported in
Ref. \protect\cite{AS96} for rolls and hexagons respectively. The solid
line is the stability boundary for straight rolls in the laterally infinite
system.}
\label{balloon}
\end{figure}

\begin{figure}[p]
\caption{The time evolution of the patterns after changing $\epsilon$
suddenly from $\epsilon _0 = -1$ to $\epsilon _1 > 0$. Images a) to d) are
for $\Gamma = 45$ and $\epsilon _1 = 2.0$ with a top temperature of 16.5
$^\circ$C and $\Delta T_c = 4.50$ $^\circ$C. In units of $t_v = 10.5$ s, the
elapsed times since the jump in $\epsilon$ are a) t = 2, b) t = 400, c) t =
2000, and d) t = 8000. Images e to h are for $\Gamma = 55$ and $\epsilon _1 =
1.0$ with a top temperature of 19.0 $^\circ$C and $\Delta T_c = 8.56$
$^\circ$C. In units of $t_v = 6.8$ s, the elapsed times since the jump in
$\epsilon$ are e) t = 2, f) t = 600, g) t = 6000, and h) t = 10000. In both
examples the pattern in the cell interior developed from random convective
flows, and the spiral/target structures coarsened and healed to a single
spiral and some defects along the wall over a few horizontal diffusion times
$\Gamma^2 t_v$.}
\label{spirals}
\end{figure}

\begin{figure}[p]
\caption{The long-time evolution of the patterns after changing $\epsilon$
suddenly from $\epsilon _0 = -1$ to $\epsilon _1 = 1.6$ for $\Gamma =
45$. For this run, the top temperature was 18.0 $^\circ$C, and $\Delta T_c =
4.53$ $^\circ$C. The elapsed times (in units of $t_v$) since the jump in
$\epsilon$ are a) t = 7500, b) t = 19000, c) t = 30000, and d) t = 38000.}
\label{long_time}
\end{figure}







\begin{references}
\bibitem{CH93} M.~C. Cross and P.~C. Hohenberg, Rev. Mod. Phys. {\bf 65}, 851
(1993).
\bibitem{MBCA93} S.~W. Morris, E. Bodenschatz, D.~S. Cannell, and G. Ahlers,
Phys. Rev. Lett.  {\bf 71}, 2026 (1993).
\bibitem{Models} H.-W. Xi, J.~D. Gunton, and J. Vinals, Phys. Rev. Lett. {\bf
71}, 2030 (1993); M.~C. Cross and Y. Tu, Phys. Rev. Lett. {\bf 75}, 834
(1995).
\bibitem{DPW94} W. Decker, W. Pesch, and A. Weber, Phys. Rev. Lett. {\bf 73},
648 (1994).
\bibitem{Busse} F.~H. Busse and R.~M. Clever, J. Fluid Mech. {\bf 91}, 319
(1979); F.~H. Busse, in {\em Hydrodynamic Instabilities and the Transition to
Turbulence}, edited by H. Swinney and J.~P. Gollub (Springer-Verlag, Berlin,
1981), p.\ 97
\bibitem{LA96} J. Liu and G. Ahlers, Phys. Rev. Lett. {\bf 77}, 3126 (1996).
\bibitem{WAC95} M. Wu, G. Ahlers, and D.~S. Cannell, Phys. Rev. Lett. {\bf
75}, 1743 (1995).
\bibitem{AS94} M. Assenheimer and V. Steinberg, Nature {\bf 367}, 345 (1994).
\bibitem{AS96} M. Assenheimer and V. Steinberg, Phys. Rev. Lett. {\bf 76},
756 (1996).
\bibitem{LCA95} M.~D. Dominguez-Lerma, G. Ahlers and D.~S. Cannell,
Phys. Rev. E {\bf 52}, 6159 (1995).
\bibitem{Bu67}F. Busse, J. Fluid Mech. {\bf 30}, 625 (1967).
\bibitem{BBMTHCA96} J.~R. de~Bruyn, {E.~Bodenschatz, S.~W.~Morris,
S.~Trainoff, Y.~Hu}, D.~S.  Cannell, and G. Ahlers, Rev. Sci. Instrum. {\bf
67}, 2043 (1996).
\bibitem{Cr89} V. Croquette, Contemp. Phys. {\bf 30}, 153 (1989).
\bibitem{HEA95b} Y. Hu, R.~E. Ecke, and G. Ahlers, Phys. Rev. E {\bf 51},
3263 (1995).
\bibitem{ASthesis} M. Assenheimer, Ph. D thesis, The Weizmann Institute of 
Science, 1994.
\end{references}
\end{document}